\documentclass[final,5p,times]{elsarticle}

\usepackage{amssymb}
\usepackage{graphicx}
\usepackage{bm}

\journal{Physics Letters B}

\begin{document}

\begin{frontmatter}

\title{Relativistic description of BCS-BEC crossover in nuclear matter}

\author[1,2]{Bao Yuan Sun}
\author[2,1]{Hiroshi Toki\corref{corr1}}\ead{toki@rcnp.osaka-u.ac.jp}
\author[3,1,4,5,6]{Jie Meng\corref{corr1}}\ead{mengj@pku.edu.cn}
\cortext[corr1]{Corresponding author}
\address[1]{School of Physics and State Key Laboratory of Nuclear Physics and Technology, Peking University, 100871 Beijing, China}
\address[2]{Research Center for Nuclear Physics (RCNP), Osaka University, Ibaraki, Osaka 567-0047, Japan}
\address[3]{School of Physics and Nuclear Energy Engineering, Beihang University, 100191 Beijing, China}
\address[4]{Department of Physics, University of Stellenbosch, Stellenbosch, South Africa}
\address[5]{Institute of Theoretical Physics, Chinese Academy of Sciences, 100080 Beijing, China}
\address[6]{Center of Theoretical Nuclear Physics, National Laboratory of Heavy Ion Accelerator, 730000 Lanzhou, China}

\begin{abstract}
We study theoretically the di-neutron spatial correlations and the
crossover from superfluidity of neutron Cooper pairs in the $^1S_0$
pairing channel to Bose-Einstein condensation (BEC) of di-neutron
pairs for both symmetric and neutron matter in the microscopic
relativistic pairing theory.  We take the bare nucleon-nucleon
interaction Bonn-B in the particle-particle channel and the
effective interaction PK1 of the relativistic mean-field approach in
the particle-hole channel. It is found that the spatial structure of
neutron Cooper pair wave function evolves continuously from BCS-type
to BEC-type as density decreases.   We see a strong concentration of
the probability density revealed for the neutron pairs in the fairly
small relative distance around $1.5~{\rm fm}$ and the neutron Fermi
momentum $k_{Fn}\in[0.6,1.0]~{\rm fm^{-1}}$. However, from the
effective chemical potential and the quasiparticle excitation
spectrum, there is no evidence for the appearance of a true BEC
state of neutron pairs at any density.  The most BEC-like state may
appear at $k_{Fn}\thicksim0.2~{\rm fm^{-1}}$ by examining the
density correlation function. From the coherence length and the
probability distribution of neutron Cooper pairs as well as the
ratio between the neutron pairing gap and the kinetic energy at the
Fermi surface, some features of the BCS-BEC crossover are seen in
the density regions, $0.05~{\rm fm^{-1}}<k_{Fn}<0.7~(0.75)~{\rm
fm^{-1}}$, for the symmetric nuclear (pure neutron) matter.
\end{abstract}


\end{frontmatter}

Pairing correlations, as collective phenomena found in various
systems as liquid $^3$He, superconductors, atomic nuclei, and
ultracold atomic gases, play a crucial role in the fermion systems.
When an attractive interaction between two fermions is weak, i.e.,
the pairing gap is much smaller than the Fermi energy, partner
fermions can be treated as a delocalized Cooper pair with large
overlap as described in the Bardeen-Cooper-Schrieffer (BCS)
theory~\cite{PhysRev.108.1175}, displaying a strong correlation in
the momentum space. If an interaction is sufficiently strong,
fermion pairs become a spatially compact bosonic bound state and
undergo Bose-Einstein condensation (BEC), showing a strong
correlation in the coordinate space. As the strength of attractive interaction between two fermions increases, the pairing phenomenon evolves from BCS to BEC state. The transition between them, often referred as BCS-BEC crossover, brings a new insight into the pairing phenomenon.   Although the BCS and BEC limits are physically quite different, the evolution between them was found to be smooth and continuous~\cite{Leggett1980, Nozieres1985, PhysRevLett.89.130402}.

In nuclear physics, neutron pairs are expected to be strongly
correlated at some density. There are rather general reasons to
expect that di-neutron correlations in low-density nuclear matter
should be significant. It is well known that the bare
nucleon-nucleon interaction in the $^1S_0$ channel leads to a
virtual state around zero energy characterized by the large negative
scattering length $a\approx-18.5\pm0.4~{\rm
fm}$~\cite{PhysRevC.36.691}. This implies a very strong attraction
between two neutrons in the spin singlet state, although the
neutron-neutron interaction is not strong enough to form a two-body
bound state in free space. Furthermore, theoretical predictions
suggest that at around $1/10$ of the nuclear saturation density
$\rho_0$, the $^1S_0$ pairing gap may take a considerably larger
value than that around $\rho_0$~\cite{BCSBaldo1990, Takatsuka1993}.

Weakly bound neutron-rich nuclei provide optimum circumstance to
study di-neutron correlations in the low-density region, where the
density around surface is unsaturated and the couplings to the
continuum spectra play an essential role~\cite{bertsch1991ap,
meng96prl,meng98plb,meng98prl,myo02}. Originally the possible
existence of a di-neutron near the surface of nuclei was predicted
by Migdal in 1972~\cite{Migdal1972}. In his work, it was shown that
although di-neutron is unbound in vacuum, it may form a bound state
in a potential well if there is a single-particle level with energy
close to zero. Subsequently, plenty of theoretical investigation
based on either schematic or microscopic models displayed a clear
signature of the strong di-neutron correlation in the ground state,
not only in light closed-shell core plus $2n$ or $4n$
nuclei~\cite{HanJon1987, bertsch1991ap, Zhukov1993,
PhysRevLett.82.4996, barranco2001, myo02, hagino:044321, hagino:054317},
but also in medium or heavy neutron-rich nuclei~\cite{Catara1984,
Tischler98, Matsuo064326, pillet:024310}, as well as in infinite
nuclear matter~\cite{PhysRevC.56.2332, serra01}. It has been
revealed that the concentration of small sized Cooper pairs in the
surface of superfluid nuclei is a quite generic
feature~\cite{pillet:024310}. In fact, such a di-neutron correlation
is in connection with the BEC-like behavior in infinite nuclear
matter at low densities. Recently, a strong low-lying dipole
strength distribution of the two-neutron halo nucleus $^{11}$Li has
been observed experimentally~\cite{nakamura:252502}. The spectrum
was reproduced well by a three-body model with a strong di-neutron
correlation \cite{esbensen:024302, myo07}.

Theoretical and experimental progress on di-neutron correlations in
weakly bound nuclei has stimulated lots of renewed interests in
possible BCS-BEC crossover of di-neutron pairs. The systematical
analysis was performed in nuclear matter by using the bare force
given by a superposition of three Gaussian functions, the
finite-range Gogny interaction, and the zero-range contact
interaction without or with medium polarization
effects~\cite{matsuo:044309, margueron:064316, SjMao2009}. It was
shown that the correlations between two neutrons get large as
density decreases, and the spatial structure of neutron Cooper pairs
changes. As a result, the BCS-BEC crossover occurs in the uniform
matter at low densities. In particular, for the screened pairing
interaction, a di-neutron BEC state is formed in symmetric matter at
around $0.001\rho_0$. The above results have been confirmed by a
further study, where the change of sign of the density correlation
function at low momentum transfer is taken as a criterion of the
BCS-BEC crossover~\cite{Isayev:014306}. In addition, the coexistence
of BCS- and BEC-like spatial structures of neutron pairs was studied
in the halo nucleus $^{11}\rm{Li}$ as well~\cite{hagino:022506}. It
has been shown that as the distance between the center of mass of
the two neutrons and the core increases, the two-particle wave
function changes from the weak coupling BCS regime to the strongly
correlated BEC regime. This result clearly confirms that a strong
di-neutron correlation between the valence neutrons is present on
the surface of the nucleus.

Details of the effective nucleon-nucleon force in nuclei are as yet
not completely clarified since most of the experimental data in
nuclear spectroscopy are not very sensitive to the details of the
nucleon-nucleon force. Hence, in most of investigations on nuclear
pairing correlations, for convenience, the effective nuclear forces
such as zero-range contact forces or finite-range Gogny force, are
used in the particle-particle ($pp$) channel. The effective interactions in relativistic mean-field (RMF) theory are also used in the $pp$ channel~\cite{kucharek1991zpa,JunLiBCS}. In order to obtain reasonable values for the gap parameter, one has to introduce
an effective factor~\cite{JunLiBCS}. Hence, the properties of Cooper
pairs and BCS-BEC crossover from these existing phenomenological
calculations need further check. Furthermore, in the low-density
limit, the interaction in the $pp$ channel approaches the bare
nucleon-nucleon interaction. Therefore, it is interesting to use the
realistic bare nucleon-nucleon interaction in the $pp$ channel and
explore the BCS-BEC crossover at low densities.

In this work, the di-neutron spatial correlations in the $^1S_0$
channel will be studied in a relativistic pairing theory, i.e.,
relativistic Hartree-Fock-Bogoliubov (RHFB) theory~\cite{kucharek1991zpa}, with the realistic bare nucleon-nucleon interaction, i.e., the relativistic Bonn potential~\cite{Machleidt}, in the $pp$ channel. Then the BCS-BEC crossover phenomenon and the possibility of di-neutron BEC states at low densities will be discussed.

The RMF theory~\cite{Serot:1986} has attracted lots of attentions
during the last decades due to its great success in describing both
nuclear matter and finite nuclei near or far from the stability
line~\cite{Ring:1996, Meng:2006}. In the RHFB theory, meson fields
are treated dynamically beyond the mean-field theory to provide the
pairing field via the anomalous Green's
functions~\cite{kucharek1991zpa}. In the case of infinite
nuclear matter, the Dirac-Hartree-Fock-Bogoliubov equation reduces
to the usual BCS equation. For the $^1S_0$ channel, the pairing gap
function $\Delta(p)$ is,
 \begin{equation}\label{Eq:GapEq0}
   \Delta(p) = -\frac{1}{4\pi^2}\int_0^\infty v_{pp}(k,p)\frac{\Delta(k)}{2E_k}k^2dk,
 \end{equation}
where $v_{pp}(k,p)$ is the matrix elements of nucleon-nucleon
interaction in the momentum space for the $^1S_0$ pairing channel,
and $E_k$ is the quasi-particle energy,
\begin{equation}
 E_k = \sqrt{(\varepsilon_k - \mu)^2 + \Delta(k)^2},
\end{equation}
with the single-particle energy $\varepsilon_k$, and the chemical
potential $\mu$. The corresponding normal and anomalous density
distribution function have the form,
 \begin{equation}\label{Eq:rhokappa}
   \rho_k = \frac{1}{2} \left[ 1 - \frac{\varepsilon_k - \mu}{E_k} \right],\quad
   \kappa_k = \frac{\Delta(k)}{2E_k}.
 \end{equation}

The single-particle energy $\varepsilon_k$ follows from the standard
RMF approach~\cite{Serot:1986},
 \begin{equation}
 \varepsilon_k = \Sigma_0 + \sqrt{k^2 + M^{\ast2}},
 \end{equation}
where the scalar mass $M^\ast=M+\Sigma_S$, $\Sigma_0$  the vector
potential and $\Sigma_S$ the scalar potential. For nuclear matter
with given baryonic density $\rho_b$ and isospin asymmetry
$\zeta=(\rho_n-\rho_p)/\rho_b$, the above equations can be solved by
a self-consistent iteration method with no-sea approximation.

The relativistic Bonn potential is used in the $pp$ channel, which
has a proper momentum behavior determined by the scattering data up to high energies~\cite{Machleidt}. It is defined as the sum of
one-boson-exchange (OBE) of the several mesons $\phi=\sigma, \omega,
\pi, \rho, \eta, \delta$. The matrix elements
$v_{pp}(\bm{k},\bm{p})$ is
 \begin{equation}
  v_{pp}(\bm{k},\bm{p}) =
  \sum_\phi\frac{\eta_\phi}{2\varepsilon^\ast_k\varepsilon^\ast_p}A_\phi(\bm{k},\bm{p})D_\phi(\bm{q}^2)F_\phi^2(\bm{q}^2),
 \end{equation}
where $\varepsilon_k^\ast$ is the effective single-particle energy
 \begin{equation}
 \varepsilon_k^\ast = \sqrt{k^2 + M^{\ast2}}.
 \end{equation}
$D_\phi(\bm{q}^2)$ is a meson propagator with momentum transfer $\bm{q}=\bm{k}-\bm{p}$, and $F_\phi(\bm{q}^2)$ is a form factor of monopole type,
 \begin{equation}
  D_\phi(\bm{q}^2) = \frac{1}{\bm{q}^2+m_\phi^2},\quad
  F_\phi(\bm{q}^2) = \frac{\Lambda_\phi^2 - m_\phi^2}{\bm{q}^2 + \Lambda_\phi^2}.
 \end{equation}
with $m_\phi$ the mass of meson, and $\Lambda_\phi$ the cutoff
parameter. $\eta_\phi$ and $A_\phi(\bm{k},\bm{p})$ are the vertex
functions of the relativistic Bonn potential. For the $^1S_0$
pairing channel, the matrix element $v_{pp}(k,p)$ is just the
average of $v_{pp}(\bm{k}, \bm{p})$ over the angle $\theta$ between
the vectors $\bm{k}$ and $\bm{p}$,
 \begin{equation}
 v_{pp}(k,p) = \int_0^\pi v_{pp}(\bm{k}, \bm{p})\sin\theta d\theta.
 \end{equation}
Bonn-B potential~\cite{Machleidt} is adopted for $v_{pp}(k,p)$.
For the mean-field calculation in the particle-hole ($ph$) channel,
the effective interaction PK1~\cite{Long04} is used, since the results
do not depend sensitively on various other parameter sets~\cite{sugahara94}.

\begin{figure}[h]
 \centering
 \includegraphics[width=0.4\textwidth]{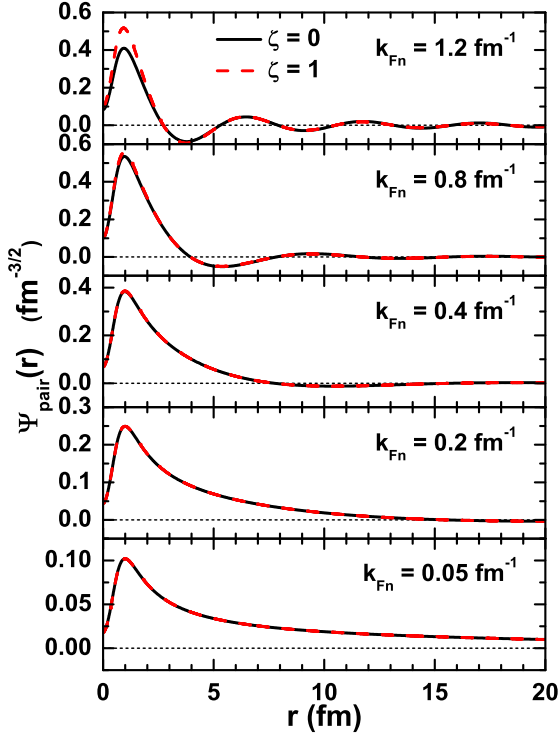}
 \caption{Neutron Cooper pair wave function in
coordinate space $\Psi_{\rm pair}(r)$ as a function of the relative
distance $r$ between the pair partners at several neutron Fermi
momenta $k_{Fn}$ in symmetric nuclear matter ($\zeta=0$, black solid
lines) and pure neutron matter ($\zeta=1$, red dashed lines). Notice
the different scales on the ordinate in the various panels.}
 \label{Fig1}
\end{figure}

To investigate the spatial structure of neutron Cooper pairs, it is
useful to look into its wave function represented as a function of
the relative distance $r=\left|\bm{r}_1-\bm{r}_2\right|$ between the
pair partners. The Cooper pair wave function in momentum space
$\Psi_{\rm pair}(k)$ is just the anomalous density $\kappa_k$, and
its coordinate representation is deduced from the Fourier transform,
 \begin{equation}
 \Psi_{\rm pair}(r) = \frac{C}{(2\pi)^3}\int
 \kappa_ke^{i\bm{k}\cdot\bm{r}}d\bm{k},
 \end{equation}
where $C$ is the constant determined from the normalization
condition
\begin{equation}
 \int \left|\Psi_{\rm pair}(r)\right|^2 r^2 dr=1.
\end{equation}

In Fig.~\ref{Fig1}, the wave function of the neutron Cooper pairs
$\Psi_{\rm pair}(r)$ is shown as a function of the relative distance
$r$ between the pair partners at different neutron Fermi momenta
$k_{Fn}$. Nearly identical results are given for the symmetric
nuclear matter ($\zeta=0$) and pure neutron matter ($\zeta=1$),
except for $k_{Fn}=1.2~{\rm fm^{-1}}$, where a larger amplitude of
the first peak is obtained in pure neutron matter.

The radial shape of $\Psi_{\rm pair}(r)$ changes as density
decreases. When $k_{Fn}=1.2~{\rm fm^{-1}}$ and $0.8~{\rm fm^{-1}}$,
the profile of the wave function $\Psi_{\rm pair}(r)$ shows an
oscillatory behavior convoluted by a decreasing exponent, which is a
typical behavior of BCS state. A significant amplitude outside the
first node is observed. As density goes down to $k_{Fn}=0.4~{\rm
fm^{-1}}$ and $0.2~{\rm fm^{-1}}$, the wave function becomes compact in shape and the oscillation disappears, resembling the strong coupling BEC-like bound state. This indicates that a possible
BCS-BEC crossover may occur in uniform matter at such low densities.  At very dilute density of $k_{Fn}=0.05~{\rm fm^{-1}}$, the wave function becomes more extended again and approaches to zero slowly.

Turning ones attention to short distance $r$, a suppressed amplitude
around $r=0~{\rm fm}$ is displayed at all densities, that is
attributed to the strong repulsive contribution of $v_{pp}(k,p)$ at
high momentum. In addition, for all the densities a peak appears
around $r=1.0~{\rm fm}$, around which it has been confirmed that the
pairing potential in the coordinate space $v_{pp}(r)$ reaches the
strongest attraction~\cite{serra01}.

Furthermore, it is noteworthy that the evolution of the wave
function with the density in Fig.~\ref{Fig1} is similar to those for
the proton-neutron pairing in ref.~\cite{Baldo975} in which a true
bound state of deuterons is predicted in the low-density limit.
Thus, one needs to explore whether a true di-neutron BEC bound state appears in nuclear matter as well.

\begin{figure}[h]
 \centering
 \includegraphics[width=0.45\textwidth]{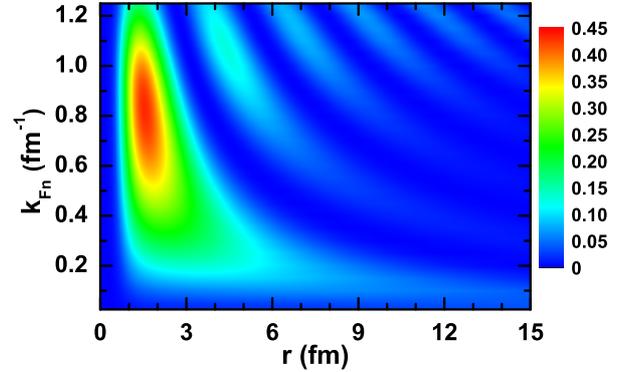}
 \caption{A two-dimensional plot for the probability
density $r^2|\Psi_{\rm pair}(r)|^2$ of the neutron Cooper pairs as a
function of the neutron Fermi momentum $k_{Fn}$ and the relative
distance $r$ between the pair partners in symmetric nuclear matter.}
 \label{Fig2}
\end{figure}

To illustrate a clear picture of di-neutron correlations, a
two-dimensional plot for the probability density $r^2|\Psi_{\rm
pair}(r)|^2$ of the neutron Cooper pairs for symmetric nuclear
matter is shown in Fig.~\ref{Fig2}, as a function of the neutron
Fermi momentum $k_{Fn}$ and the relative distance $r$ between the
pair partners. A strong concentration of the probability density is
revealed for the pair partners in a surprisingly small relative
distance of $r\thicksim1.5~{\rm fm}$ in the density region about
$0.6~{\rm fm^{-1}}<k_{Fn}<1.0~{\rm fm^{-1}}$. This behavior has the
same physical essence as the di-neutron configuration in the surface
of superfluid nuclei as discussed in Refs.~\cite{Matsuo064326,
pillet:024310, hagino:022506}. As a matter of fact, the pairing gap
at the Fermi surface $\Delta_{Fn}\equiv\Delta(k_{Fn})$ also depends
strongly on $k_{Fn}$, and achieves a large value together with
probability density in such a density region.

When the Fermi momentum is close to the saturation density, the
probability density $r^2|\Psi_{\rm pair}(r)|^2$ indicates a
considerable distribution outside the first peak and extends to a
large distance $r$. This phenomenon is well understood by the orthogonalization of the wave function for the paired particles with the remaining particles due to the Pauli principle, which expels the Cooper pair wave function outside. As density decreases, these amplitudes die
out gradually and a compact structure appears at about $0.2~{\rm
fm^{-1}}<k_{Fn}<0.5~{\rm fm^{-1}}$. While going to lower densities
of $k_{Fn}<0.2~{\rm fm^{-1}}$, the probability density has a very
small value and exhibits a long tail at large relative distance $r$.
This behavior may be responsible for the appearance of halo
structure in several neutron-rich nuclei at far distance away from
the center~\cite{meng96prl,meng98plb,myo07}.

Since a strong di-neutron correlation is revealed at low densities
from the Cooper pair wave function and the corresponding probability
density, it is deserved to study the possibility of Bose-Einstein
condensation and the BCS-BEC crossover phenomenon of di-neutrons as
density decreases.

\begin{figure}[h]
 \centering
 \includegraphics[width=0.4\textwidth]{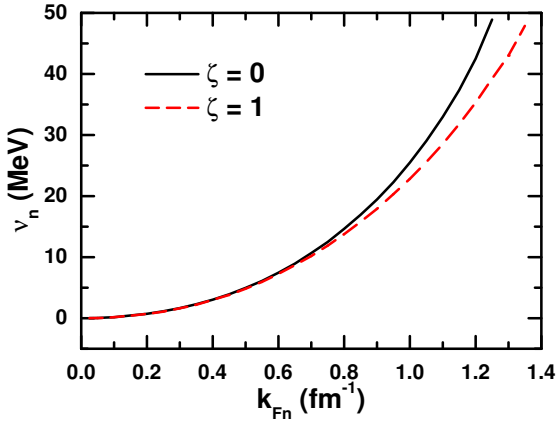}
 \caption{Effective neutron chemical potential $\nu_n$ as a function of the
neutron Fermi momentum $k_{Fn}$ in symmetric nuclear matter (black
solid line) and pure neutron matter (red dashed line).}
 \label{Fig3}
\end{figure}

In the whole debate around BCS-BEC transition the chemical potential
$\mu$ plays a central role. It has been proved in an easy way that
the gap in Eq.~(\ref{Eq:GapEq0}) together with the normal and
anomalous density distribution function in Eq.~(\ref{Eq:rhokappa})
goes over into the Schr\"{o}dinger-like equation for the neutron
pair wave function $\Psi_{\rm pair}(k)$, and the corresponding
energy eigenvalue is related to the chemical
potential~\cite{Nozieres1985}. In the case of the RMF approach, it
is expressed as
 \begin{eqnarray}\label{Eq:SchR}
 2e(k)\Psi_{\rm pair}(k) + \frac{(1-2\rho_k)}{4\pi^2}\int_0^\infty v_{pp}(k,p)p^2dp\Psi_{\rm pair}(p)&&\nonumber\\
 = 2\nu_n\Psi_{\rm pair}(k),&&
 \end{eqnarray}
where $e(k)$ is the neutron kinetic energy,
 \begin{equation}
 e(k)=\sqrt{k^2 + M^{\ast2}} -M^\ast.
 \end{equation}
$\nu_n$ is the effective neutron chemical potential obtained by
deducting momentum independent part from the chemical potential
$\mu_n$,
 \begin{equation}
 \nu_n=\mu_n - \Sigma_0 -M^\ast,
 \end{equation}
which could be regarded as half of the ``binding energy'' of a
Cooper pair. At transition from the BCS regime of large overlapping
neutron Cooper pairs to the BEC regime with true di-neutron bound
states, $\nu_n$ is supposed to turn from positive to negative
values.

In Fig.~\ref{Fig3}, the effective neutron chemical potential $\nu_n$
as a function of the Fermi momentum $k_{Fn}$ in nuclear matter is
given. The values of $\nu_n$ decrease monotonically in both
symmetric nuclear matter and pure neutron matter as density goes
down, and come very close to zero at dilute area but never turn
negative. It is then expected in this case that a true di-neutron
bound state does not occur in nuclear matter, but the transition to
BEC is very close in the low-density limit.

\begin{figure}[h]
 \centering
 \includegraphics[width=0.4\textwidth]{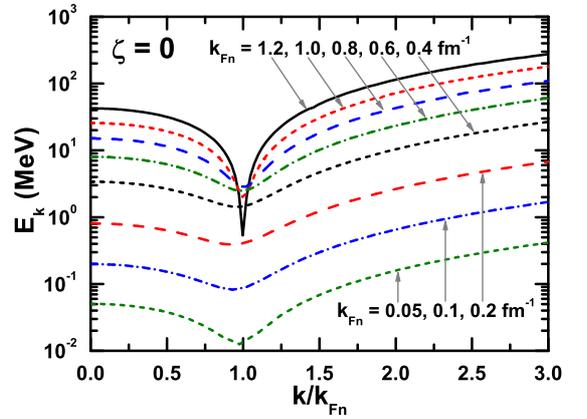}
 \caption{Neutron quasiparticle excitation energy
$E_k$ as a function of the ratio of the neutron momentum to the
Fermi momentum $k/k_{Fn}$ at different neutron Fermi momenta
$k_{Fn}$ in symmetric nuclear matter.}
 \label{Fig4}
\end{figure}

The qualitative changes in the quasiparticle excitation spectrum
have been suggested as another clear distinction between the BCS and
BEC limits~\cite{Parish2005}, i.e., the momentum corresponding to
the minimum in the excitation spectrum will shift from finite
momentum in the BCS limit to zero momentum in the BEC limit, and the
value of the pairing gap also changes from $\Delta$ in the BCS limit
to $\sqrt{\mu^2+\Delta^2}$ in the BEC limit.

In Fig.~\ref{Fig4}, the quasiparticle excitation spectra $E_k$ is
shown as functions of $k/k_{Fn}$ at several neutron Fermi momenta in
symmetric nuclear matter. The minimum of $E_k$ always appears around
the Fermi momentum $k_{Fn}$ for all the densities, i.e., the minimum
of $E_k$ approaches to zero momentum with the corresponding Fermi
momentum but become zero only when $k_{Fn}=0$. Similar results are
obtained for the pure neutron matter as well. Therefore, from the
quasiparticle excitation spectrum, there is no evidence for the
appearance of a true BEC bound state of neutron pairing with the
realistic bare nucleon-nucleon interaction Bonn-B at any density, in
agreement with that from the effective chemical potential $\nu_n$.

\begin{figure}[h]
 \centering
 \includegraphics[width=0.4\textwidth]{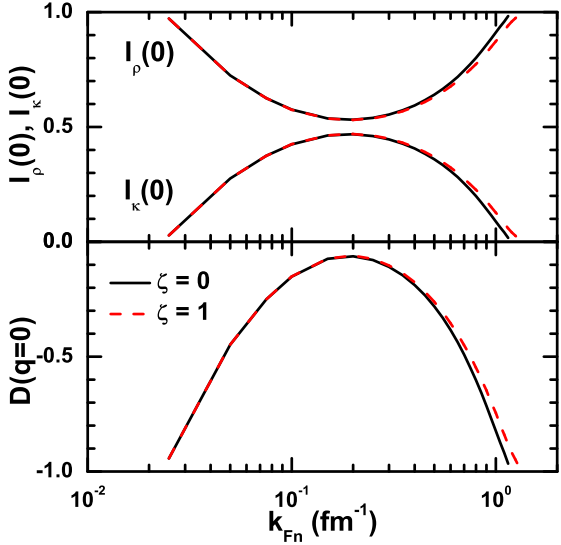}
 \caption{Zero-momentum transfer density correlation
function $D(q=0)$ and its normal and anomalous density
contributions, $I_\rho(0)$ and $I_\kappa(0)$, as a function of the
neutron Fermi momentum $k_{Fn}$ in symmetric nuclear matter (black
solid lines) and pure neutron matter (red dashed lines).}
 \label{Fig5}
\end{figure}

The analogies with other low-density Fermi systems, such as
ultracold atomic gases, could be used to study the BCS-BEC crossover in nuclear matter as well. The neutron density correlation function is such a quantity that gives prominence to the relative strength between the mean field and the pairing field, which is defined
as~\cite{mihaila:090402, Isayev:014306},
 \begin{equation}
 D(q)=I_\kappa(q)- I_\rho(q).
 \end{equation}
At zero momentum transfer, the normal and anomalous density
contributions $I_\rho$ and $I_\kappa$ respectively read
 \begin{eqnarray}
 I_\rho(0) &=&\frac{1}{\pi^2\rho_n}\int_0^\infty \rho_k^2k^2dk,\label{I_0}\\
 I_\kappa(0) &=&\frac{1}{\pi^2\rho_n}\int_0^\infty \kappa_k^2k^2dk,
 \end{eqnarray}
and they satisfy the sum rule
 \begin{equation}
 I_\rho(0) +I_\kappa(0) = 1.
 \end{equation}

The change of sign of the density correlation function at low
momentum transfer is taken as a criterion of the BCS-BEC
crossover~\cite{mihaila:090402}, although further study argued that
this criterion can be trusted only at small isospin
asymmetry~\cite{Isayev:014306}.

In Fig.~\ref{Fig5}, the density correlation functions at
zero-momentum transfer $D(0)$ and its normal and anomalous density
contributions, $I_\rho(0)$ and $I_\kappa(0)$, are shown as functions
of $k_{Fn}$ in nuclear matter. With decreasing density, $I_\rho(0)$
first drops down from $1.0$, reaches a minimum at
$k_{Fn}\thicksim0.2~{\rm fm^{-1}}$
($\rho_n/\rho_0\thicksim10^{-3}$), and then increases to $1.0$
again. While $I_\kappa(0)$ gives an opposite trend and has no
contribution when approaching to either the saturation density or
the dilute density. This leads to a peak for the density correlation
function $D(0)$ around $k_{Fn}=0.2~{\rm fm^{-1}}$, where the most
BEC-like state may appear. It is consistent with the result of
$\Psi_{\rm pair}(r)$ shown in Fig.~\ref{Fig1}, where the wave
function behaves like a spatially compact bound state at
$k_{Fn}=0.2~{\rm fm^{-1}}$. Because the anomalous density
contribution $I_\kappa(0)$ is always smaller than the normal one
$I_\rho(0)$, the density correlation function never changes sign at
all. According to the criterion mentioned above, di-neutrons are
not in BEC state but just in the transition region between BCS and BEC regimes, that verifies the conclusion from the effective chemical
potential and the quasiparticle excitation spectrum.

\begin{figure}[h]
 \centering
 \includegraphics[width=0.45\textwidth]{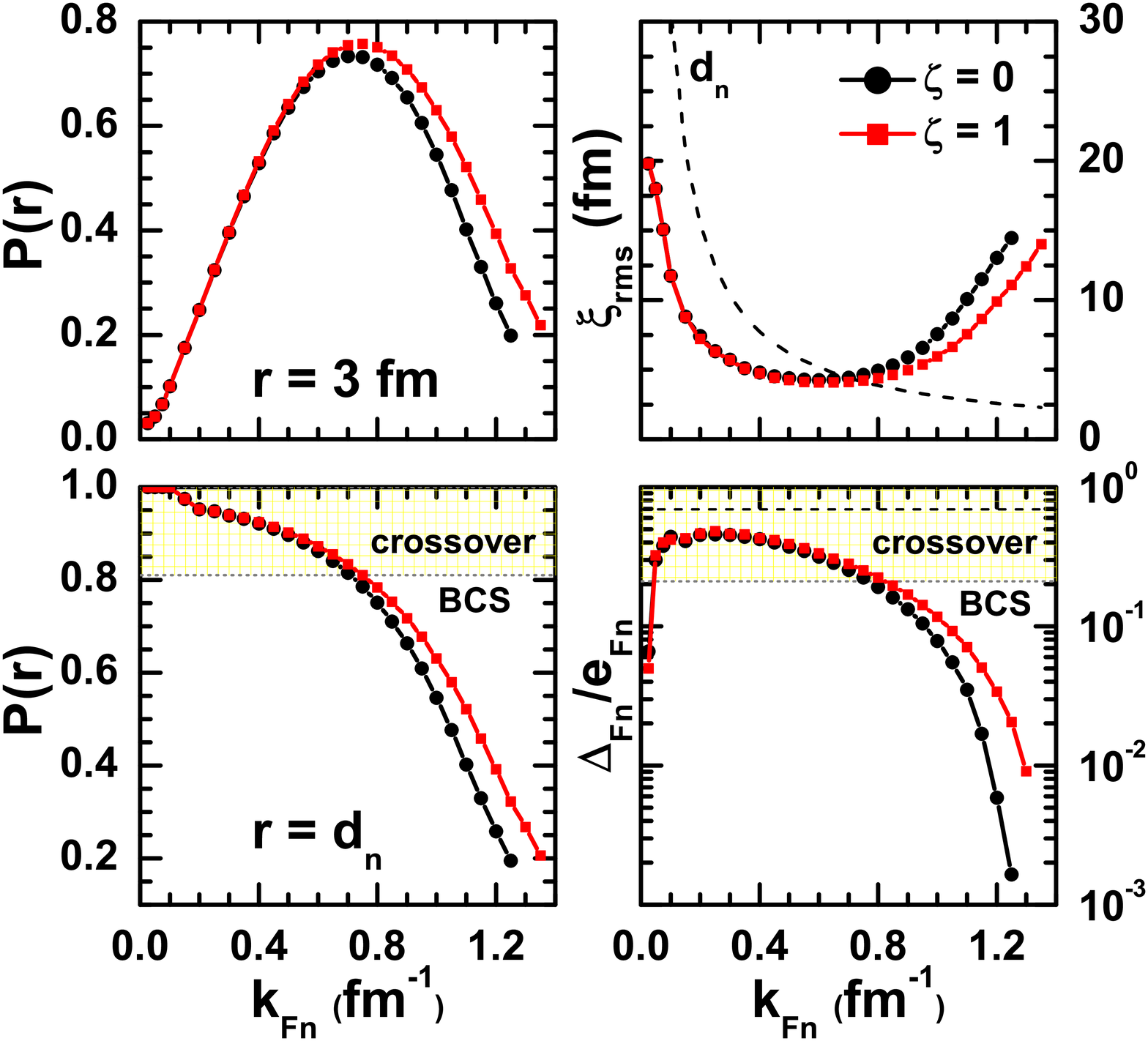}
 \caption{Probability $P(r)$ for the partner neutrons
correlated within $r=3~{\rm fm}$  (upper-left panel) and the average
inter-neutron distance $r=d_n$ (lower-left panel) as a function of
the neutron Fermi momentum $k_{Fn}$. The black line with filled
circle is for the symmetric nuclear matter and the red line with
filled square is for the pure neutron matter. The corresponding rms
radius $\xi_{\rm rms}$ of the neutron Cooper pair is plotted in
upper-right panel in comparison with the average inter-neutron
distance $d_n$ (the dashed curve). The ratio $\Delta_{Fn}/e_{Fn}$
between the neutron pairing gap at the Fermi surface $\Delta_{Fn}$
and the neutron Fermi kinetic energy $e_{Fn}$ as a function of
$k_{Fn}$ is plotted in lower-right panel. The referred BCS-BEC
crossover regions (yellow grid) and the unitary limit (dashed line)
from the regularized pairing model~\cite{matsuo:044309,
margueron:064316} for $P(d_n)$ and $\Delta_{Fn}/e_{Fn}$ are given
respectively.}
 \label{Fig6}
\end{figure}

To describe the BCS-BEC crossover quantitatively, lots of
characteristic quantities have been introduced~\cite{matsuo:044309,
margueron:064316}. The rms radius of the neutron Cooper pair
$\xi_{\rm rms}$, i.e., coherence length, is a straightforward
measure to evaluate the pairing size,
 \begin{equation}
 \xi_{\rm rms}^2=\langle r^2\rangle
 = \frac{\int \left| \Psi_{\rm pair}(r)\right|^2 r^4dr}{\int\left| \Psi_{\rm pair}(r)\right|^2 r^2dr}
 = \frac{\int_0^\infty \left( \frac{\partial}{\partial k}\kappa_k\right)^2 k^2dk}{\int_0^\infty \kappa_k^2 k^2dk},
 \end{equation}
which can be calculated either from the Cooper pair wave function
$\Psi_{\rm pair}(r)$ or from $\kappa_k$ in the momentum space. The
latter one is more suitable since the gap equation is solved in
momentum space. The coherence length $\xi_{\rm rms}$ has a solid
meaning even in the strong coupling BEC case or the crossover region
between BCS and BEC. For comparison, the average inter-neutron
distance $d_n\equiv \rho_n^{-1/3}$ is taken into account. If
$\xi_{\rm rms}>d_n$, the neutron Cooper pair is interpreted as an
extended BCS-like pair, otherwise it will be considered as a compact
BEC-like pair.

The coherence lengths $\xi_{\rm rms}$ of the neutron Cooper pairs as
a function of the neutron Fermi momentum $k_{Fn}$ in both symmetric
nuclear matter and pure neutron matter are drawn in the upper-right
panel of Fig.~\ref{Fig6}. While $k_{Fn}$ decreases, the coherence
lengths shrinks to $\sim 5~{\rm fm}$ at density of $0.4~{\rm
fm^{-1}}<k_{Fn}<0.8~{\rm fm^{-1}}$ ($0.4~{\rm
fm^{-1}}<k_{Fn}<0.9~{\rm fm^{-1}}$) for symmetric nuclear (pure
neutron) matter. Then it expands again at very low densities. These
features can be understood as follows.  In comparison with the
virtual $S$-state in the vacuum, the neutron pairs feel an
extra binding due to the pairing correlation as density increases;
while at large $k_{Fn}$, the neutron pairs will strongly increase in
size due to the Pauli principle. The difference of the coherence
lengths between $\zeta=0$ and $\zeta=1$ at high densities is caused by the difference of the scalar mass $M^\ast$. When $k_{Fn}<0.7~{\rm
fm^{-1}}$ ($k_{Fn}<0.75~{\rm fm^{-1}}$) for symmetric nuclear (pure
neutron) matter, it is seen that the coherence length $\xi_{rms}$ is
smaller than the average inter-neutron distance $d_n$, which could
be treated as the  signature of the BCS-BEC crossover.

In order to characterize the spatial correlation of the neutron
Cooper pairs, one can introduce the probability $P(r)$,
 \begin{equation}
 P(r) = \int_0^r\left| \Psi_{\rm pair}(r^\prime)\right|^2 r^{\prime 2} dr^\prime,
 \end{equation}
which gives the probability for the pair partners within a relative
distance $r$.

The probability $P(r)$ for the neutron pair partners correlated
within $r=3~{\rm fm}$ is shown in the upper-left panel of
Fig.~\ref{Fig6} as a function of $k_{Fn}$ in symmetric nuclear
matter and pure neutron matter. It reaches the maximum value of
about $0.75$ around $k_{Fn}=0.75~{\rm fm^{-1}}$, and goes down at
both lower or larger Fermi momentum. In the lower-left panel of
Fig.~\ref{Fig6}, the probability $P(d_n)$, i.e., $P(r)$ for the pair
partners within the average inter-neutron distance $r=d_n$, is
plotted as well. The values of $P(d_n)$ increase monotonically with
decreasing density, and approach to $1.0$ at dilute area. From a
regularized pairing model~\cite{PhysRevLett.71.3202}, it is
suggested that the BCS-BEC crossover is determined by
$P(d_n)>0.807$~\cite{matsuo:044309, margueron:064316}. This limits
the Fermi momentum $k_{Fn}<0.7~{\rm fm^{-1}}$ in symmetric nuclear
matter and $k_{Fn}<0.75~{\rm fm^{-1}}$ in pure neutron matter for
the realization of the BCS-BEC crossover.

The BCS-BEC crossover is also explored by the ratio
$\Delta_{Fn}/e_{Fn}$ between the neutron pairing gap at the Fermi
surface $\Delta_{Fn}$ and the neutron Fermi kinetic energy
$e_{Fn}\equiv e(k_{Fn})$. If the ratio is large enough, the neutron
pairing is expected to be in the strong coupling regime. From the
regularized model, the BCS-BEC crossover region is estimated as
$0.21<\Delta_n/e_{Fn}<1.33$ and the unitary limit is given at
$\Delta_n/e_{Fn}=0.69$~\cite{matsuo:044309, margueron:064316}.

In the lower-right panel of Fig.~\ref{Fig6}, the ratios
$\Delta_{Fn}/e_{Fn}$ plotted as a function of the neutron Fermi
momentum are shown for both symmetric nuclear matter and pure
neutron matter. As density decreases, the values grow up first and
go into BCS-BEC crossover region at $k_{Fn}\sim0.75~{\rm fm^{-1}}$
for symmetric nuclear matter and $k_{Fn}\sim0.8~{\rm fm^{-1}}$ for
pure neutron matter. It is found that the curves never intersect
with the line of the unitary limit, namely, the neutron pairing does
not go into the BEC regime. Then at very low densities with
$k_{Fn}<0.05~{\rm fm^{-1}}$, the results decrease again and return
back to the BCS region. Based upon the above analyses, the BCS-BEC
crossover is marked in the density region with $0.05~{\rm
fm^{-1}}<k_{Fn}<0.7~{\rm fm^{-1}}$ for the symmetric nuclear matter
and $0.05~{\rm fm^{-1}}<k_{Fn}<0.75~{\rm fm^{-1}}$ for the pure
neutron matter.

In conclusion, the di-neutron spatial correlations and the BCS-BEC
crossover phenomenon for nuclear matter in the $^1S_0$ channel has
been investigated based on the microscopic relativistic pairing
theory with the effective RMF interaction PK1 in the $ph$ channel
and the realistic bare nucleon-nucleon interaction Bonn-B in the
$pp$ channel. It is found that the spatial structure of neutron
Cooper pair wave function evolves continuously from BCS-type to
BEC-type as density decreases, and a strong concentration of the
probability density is revealed for the pair partners in the fairly
small relative distance around $1.5~{\rm fm}$ and the neutron Fermi
momentum $k_{Fn}\in[0.6,1.0]~{\rm fm^{-1}}$, which is correlated
with the di-neutron configuration in the surface of superfluid
nuclei. In light of the evidence from the effective chemical
potential, the quasiparticle excitation spectrum and the density
correlation function, a true di-neutron BEC bound state does not
occur at any density in nuclear matter. Neutron pairing is just in
the transition region from BCS to BEC regime at low densities, and
the most BEC-like state may appear at $k_{Fn}\thicksim0.2~{\rm
fm^{-1}}$. From several characteristic quantities, such as the
coherence length $\xi_{\rm rms}$, the probability $P(r)$ with $r=
3~{\rm fm}$ and $r=d_n$, and the ratio $\Delta_{Fn}/e_{Fn}$, the
BCS-BEC crossover is found in the density region about $0.05~{\rm
fm^{-1}}<k_{Fn}<0.7~{\rm fm^{-1}}$ for symmetric nuclear matter and
$0.05~{\rm fm^{-1}}<k_{Fn}<0.75~{\rm fm^{-1}}$ for pure neutron
matter. The results reveal a strong correlation of neutron pairs at
such densities.

\section*{Acknowledgements}
B. S. acknowledges the support of FrontierLab@OsakaU Program for his
three months stay in Osaka University, where part of this work has
been performed. This work is partly supported by the Major State
Basic Research Development Program (2007CB815000), and the National
Natural Science Foundation of China (10975008 and 10775004).

\end{document}